# High-energy, few-cycle light pulses tunable across the vacuum ultraviolet

José R. C. Andrade[1,*], Martin Kretschmar[1,*], Rostyslav Danylo[1], Stefanos Carlström[1], Tobias Witting[1], Alexandre Mermillod-Blondin[1], Serguei Patchkovskii[1], Misha Yu Ivanov[1,2], Marc J. J. Vrakking[1], Arnaud Rouzée[1] and Tamas Nagy[1,*]

[1]Max Born Institute for Nonlinear Optics and Short Pulse Spectroscopy, Max-Born-Straße 2A, 12489 Berlin, Germany

[2]Institut für Physik, Humboldt-Universität zu Berlin, Newtonstraße 15, 12489 Berlin, Germany

Contact details:

José R. C. Andrade: jose.andrade@mbi-berlin.de, +49 30 6392-1277

Martin Kretschmar: martin.kretschmar@mbi-berlin.de, +49 30 6392-1271

Rostyslav Danylo: rostyslav.danylo@mbi-berlin.de

Stefanos Carlström: stefanos.carlstrom@mbi-berlin.de

Tobias Witting: witting@mbi-berlin.de

Alexandre Mermillod-Blondin: mermillod@mbi-berlin.de

Serguei Patchkovskii: serguei.patchkovskii@mbi-berlin.de

Misha Yu Ivanov: mikhail.ivanov@mbi-berlin.de

Marc J. J. Vrakking: vrakking@mbi-berlin.de

Arnaud Rouzée: rouzee@mbi-berlin.de

Tamas Nagy: nagy@mbi-berlin.de, +49 30 6392-1270

**In the last few decades the development of ultrafast lasers has revolutionized our ability to gain insight into light-matter interactions. The appearance of few-cycle light sources available from the visible to the mid-infrared spectral range and the development of attosecond extreme ultraviolet and x-ray technologies provide for the first time the possibility to directly observe and control ultrafast electron dynamics in matter on their natural time scale. However, few-fs sources have hardly been available in the deep ultraviolet (DUV; 4-6 eV, 300-200 nm) and are unavailable in the vacuum ultraviolet (VUV; 6-12 eV, 200-100 nm) spectral range, corresponding to the photon energies required for valence excitation of atoms and molecules. Here, we generate VUV pulses with µJ energy tunable between 160 and 190 nm via resonant dispersive wave emission during soliton self-compression in a capillary. We fully characterize the pulses in situ using frequency-resolved optical gating based on two-photon photoionization in noble gases. The measurements reveal that in most of the cases the pulses are shorter than 3 fs. These findings unlock the potential to investigate ultrafast electron dynamics with a time-resolution that has been hitherto inaccessible when using VUV pulses.**

Ultrafast electron dynamics became accessible to investigations in the last two decades with the development of attosecond extreme ultraviolet (XUV) light sources[1]. However, owing to its high photon energy (above 20 eV) a single XUV photon can directly ionize matter, which limits the scope of investigations to ionization-driven or core-excited electron dynamics[2]. To answer a manifold of fundamental questions in physics and chemistry, ultraviolet, visible or infrared pulses of the shortest possible duration are needed[3]. Understanding e.g. the photoexcitation dynamics of molecules upon visible and ultraviolet absorption presents a significant challenge, particularly due to the break-down of the Born-Oppenheimer approximation according to which fast-moving electrons can instantaneously adjust to the slower motion of nuclei. In electronically excited molecules, the presence of closely spaced electronic states can result in strongly coupled electronic and nuclear dynamics along one or more reactive modes, making it impossible to treat them separately. These couplings lead to radiationless non-adiabatic transitions occurring on ultrafast timescales.

The generation of few-fs ultraviolet pulses that would allow for temporally resolving ultrafast valence electron dynamics is a very challenging task. In particular, the vicinity of electronic resonances present in virtually all materials in the UV spectral range, results in both increased linear absorption and high dispersion. The latter is detrimental to the temporal resolution smearing out the initially short pulses to long duration in the course of propagation. In practice, no material can be tolerated as a propagation medium making the handling of ultrashort UV pulses extremely difficult. Indeed, only a few groups achieved µJ-level sub-10 fs pulses in the DUV by standard nonlinear optical methods, such as four-wave mixing in a hollow-core fiber (HCF)[4,5] and achromatic second harmonic generation[6,7] or even shorter, few-cycle pulses by low-order harmonic generation in noble gases[8-12]. The latter approach proved to be especially useful, not only because it generates sufficiently short pulses for the investigation of ultrafast dynamics in matter on few-fs time scales, but also because it is based on a well-engineered technology, which is routinely used nowadays for the generation of attosecond pulses in the XUV. One drawback remains though, namely the lack of spectral tunability. In principle, this could be overcome by driving the harmonic generation process with a few-cycle parametric source broadly tunable in the infrared. However, this requires a multi-octave phase-matching bandwidth of the parametric process which remains to be demonstrated.

An alternative solution towards the generation of few-cycle pulses tunable across the entire UV is offered by fiber technology. The appearance of hollow-core photonic-crystal fibers (HC-PCF)[13,14] with engineered dispersion properties and gas as nonlinear medium paved the way to exploiting physical phenomena such as the formation of optical solitons[15,16] for a new class of ultrafast light source. Optical solitons can emerge in a medium where the sign of the optical Kerr-effect is opposite to that of the dispersion. In this case the two effects can balance each other leading to a dispersionless propagation, where the pulse shape remains unchanged, or the pulses can even become compressed accompanied by a giant spectral broadening often yielding an over-octave-spanning supercontinuum. In this specific case, if the newly generated spectra reach a region where their phase velocity matches that of the soliton, a very effective energy transfer from the soliton to this spectral region can occur. This phase matching leading to the creation of resonant dispersive wave radiation mostly happens at shorter wavelengths, e.g. in the UV. The overall dispersion of the gas-filled waveguide and thus the resonant wavelength can easily be tuned across the ultraviolet by varying the gas pressure in the core[17]. Indeed, soliton self-compression and accompanying resonant dispersive-wave (RDW) generation in the UV were theoretically predicted[18] for HC-PCFs with a negative dispersion profile and shortly after also experimentally demonstrated[19,20]. Owing to the rather small dimensions of the PCF structures, these sources are capable of delivering few-femtosecond pulses with pulse energies on the order of a few tens of nanojoules, which is often insufficient for investigating ultrafast dynamics in dilute targets such as gas-phase molecules. This limitation is largely mitigated by capillaries with moderate core diameters, which can also exhibit negative dispersion, allowing the same soliton dynamics as in PCFs but in a strongly up-scaled manner, yielding pulse energies in the multi-microjoule range[21].

Although the UV pulse generation mechanism in HCFs itself is fairly simple and robust, there are a few hurdles that remain to be overcome[22]. In certain conditions the UV pulses emerging from the waveguide are expected to be close to their Fourier transform-limited temporal shape. The passage through even a thin window, usually used for enclosing the gas within the waveguide, irreparably deteriorates the pulse shape, since chirp management capabilities are scarce in the DUV and practically non-existent in the VUV. Therefore, the output of the capillary needs to be directly injected into the vacuum apparatus in which the experiments utilizing the DUV/VUV pulses are performed. The gas used as nonlinear medium is filled at the entrance side whereas the pressure gradually drops to zero towards the output[23]. Furthermore, the RDW pulse emerges superimposed on the fundamental soliton, and needs to be spectrally separated from it.

The last, and inevitably the most difficult challenge, is the full characterization of the broadband few-cycle UV pulses having a predicted typical duration between 2 fs and 5 fs. Although the first full characterization of RDW pulses in the DUV has been recently performed, no characterization has yet been reported in the much more difficult VUV range. Sub-3 fs RDW pulses tunable from 250 nm to 350 nm were measured with a self-diffraction frequency-resolved optical gating (FROG) arrangement placed in the vacuum beamline[24] utilizing a 50 µm thick $CaF_2$ plate as the nonlinear medium. While these results exploit the full potential of all-optical pulse measurement methods, such techniques are hardly applicable for pulses centred at even shorter wavelength due to excessive dispersion. The influence of high absorption and material dispersion of the nonlinear medium on the pulse characterization can be mitigated by utilizing noble gases at low pressure possessing orders of magnitude lower density than solids, thereby leading to low dispersion. In such low-density media the nonlinearities frequently utilized in pulse characterization schemes, such as low-order harmonic generation or those using the optical Kerr-effect (self-diffraction or transient grating formation) yield non-measurable signals. An alternative is to use photoionization and detect ions as a function of the delay between two pulse replicas yielding an autocorrelation signal[10]. However, while autocorrelation measurements provide an estimation of the pulse duration assuming a specific shape, they cannot fully characterize the pulse, which requires both amplitude and phase information.

Here, we present the generation and full characterization of VUV pulses with µJ energy tuned between 160 nm and 190 nm by photoionization of an atomic gas using a sequence of two pulse replicas. We utilize a variant of the FROG technique[25] based on photoelectron spectroscopy[26] (electron-FROG or in short eFROG). Our measurements are validated by comparing the results of our experiments with *ab initio* calculations of the two-photon ionization dynamics of atoms and also with simulations of the VUV pulse generation process.

**Results**

In our experiments we generate few-fs pulses with tens of μJ energy tunable across the VUV via RDW emission in a cascaded capillary arrangement seeded by a 800 nm Ti:sapphire laser. The details are summarized in the Methods section and a sketch of the full arrangement is displayed in the Supplementary Fig. S1. First, we compress the laser pulses to ~10 fs duration by passage through a gas-filled stretched flexible hollow-core fiber[27] (SF-HCF). The compressed pulses are then coupled into a second gas-filled SF-HCF where soliton self-compression and RDW emission take place. We inject helium at the entrance of the second capillary while the output side is directly connected to a vacuum beamline yielding a gradually decreasing gas pressure along the waveguide. The VUV pulse characterization arrangement is displayed in Fig. 1. The beam emerging from the second SF-HCF is first re-collimated, and then the broadband soliton content is filtered out by Brewster-reflections off two consecutive silicon plates. The filtered RDW pulses are then measured either by a VUV spectrometer and a powermeter or sent into the eFROG apparatus for full temporal characterization. In the eFROG apparatus a pair of d-shaped mirrors splits the pulse front in half and introduces a delay between the two halves. The two half-beams are focused into a velocity-map imaging (VMI) spectrometer[28] where they are used to ionize atoms of a noble gas. While scanning the delay between the two pulse replicas, we record the velocity distribution of the photoelectrons resulting from two-photon ionization of the atomic gas using a detector consisting of a sequence of two multi-channel plates (MPC), a phosphor screen and a digital camera. A detailed description of the data acquisition and processing is given in the Supplement S3.

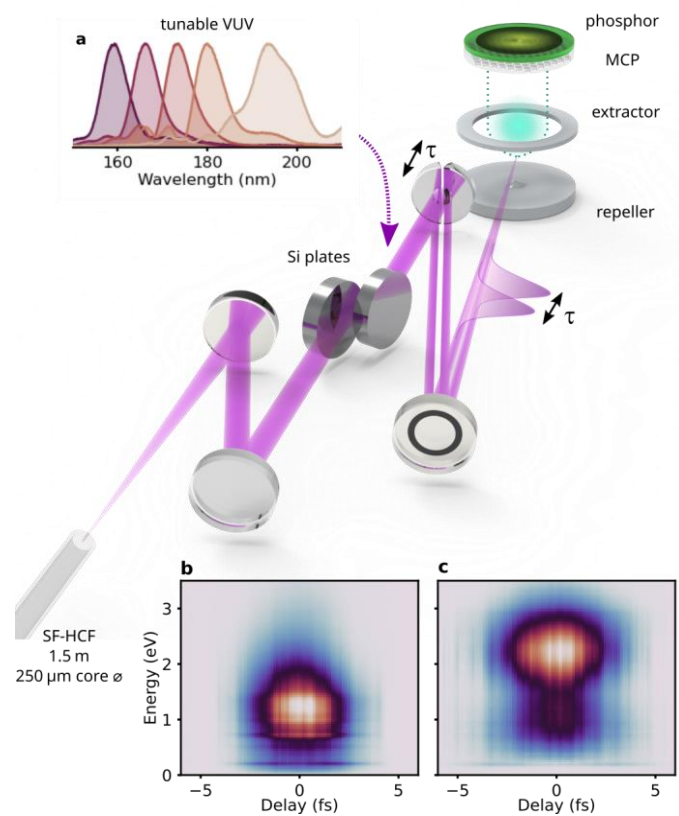

Fig. 1 Layout of the experimental arrangement used for the characterization of few-fs VUV pulses by eFROG. The RDW pulses emerging from the helium-filled capillary (SF-HCF) are first filtered by Brewster reflections off two silicon plates. Two pulse replicas are created by a split-and-delay mirror pair and then focused into a gas jet placed inside a VMI spectrometer which measures the kinetic energy spectrum of the produced photoelectrons at varying delays between the pulse replicas. The inset **a** shows measured spectra tuned across the VUV. The insets **b** and **c** display two exemplary measured eFROG traces, where xenon atoms are photoionized by VUV pulses centered at 180 nm and 170 nm, respectively. In **b** the imprint of auto-

ionizing states is observed at low energy (<1 eV) while in **c** a double-peak structure is observed due to spin-orbit splitting in the xenon atoms.

Our method to characterize the temporal profile of the VUV pulse relies on recording the delay-dependent kinetic energy distribution of photoelectrons resulting from non-resonant two-photon ionization of an atomic gas target. If the ionization process occurs on a timescale much shorter than the pulse duration, i.e. away from any resonances, this method is analogous to an all-optical second-harmonic generation (SHG) FROG, where the nonlinearity can be simply calculated in the time domain by a multiplication of the electric field by its time shifted replica. The ionization process poses clear conditions for the choice of the gas species, depending on the generated VUV spectrum: (i) two photons should be able to ionize the atoms, meaning that the photon energy of the red side of the spectrum should exceed half of the ionization potential. (ii) electronic excitation of the atoms should be excluded because it would lead to a non-instantaneous sequential ionization. Condition (ii) implies that the photon energy at the short-wavelength wing of the spectrum should remain below the lowest-lying Rydberg transition. According to these criteria the entire VUV spectral range can be covered by noble gases, as shown in the Supplementary Table S1.

Fig. 1.**a** shows VUV spectra of the pulses that were characterized, while Fig. 1.**b** and **c** represent two eFROG traces measured using pulses centered at 180 nm and 170 nm for the photoionization of xenon atoms, respectively. Both traces possess features which are unknown from SHG FROG traces: sharp horizontal lines in Fig. 1.**b** and a double-peaked structure in **c**, although the pulse spectrum itself exhibits a single peak. In order to understand the origin of these features we need to have a closer look at the atomic structure of the nonlinear medium. Rare gas atoms have two ionization thresholds close in energy that are separated by spin-orbit splitting (see Fig. 2.**a**). Due to the two different ionization channels and depending on whether the ion is formed in the $5p^5$ $^2P_{3/2}$ or the $5p^5$ $^2P_{1/2}$ state, the kinetic energy spectrum of the photoelectrons consists of two slightly shifted replicas (see Fig. 2.**c**). In case of the ionization channel leading to the formation of the ion in the $5p^5$ $^2P_{1/2}$ state, which require higher photon energies, the broad VUV spectrum might violate (i), leading to the excitation of auto-ionizing states (AIS) which modulate the low-energy region of the kinetic energy spectrum (see Fig. 2.**b**).

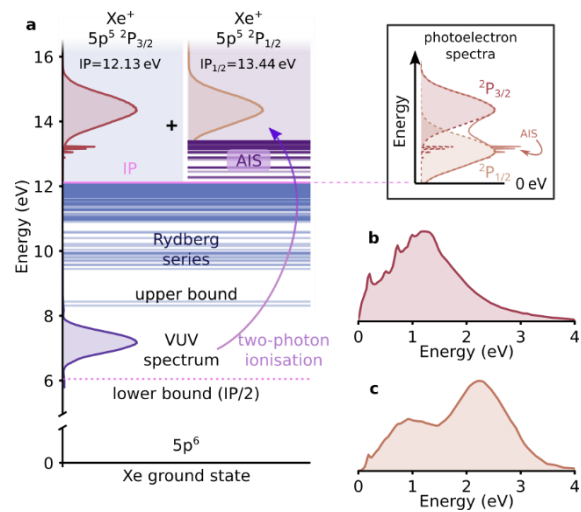

Fig. 2 Photoelectron kinetic energy spectra due to two-photon ionization of xenon. **a**, Energy level diagram of xenon showing the two possible ionization channels and the lower and upper bounds for the spectrum of the VUV pulse that can be characterized by xenon. The schematics of the resulting kinetic energy spectrum is shown on top right. Ionization producing the $5p^5$ $^2P_{3/2}$ state can occur either through direct non-resonant two-photon ionization or through two-photon excitation of an auto-ionizing state converging on the $5p^5$ $^2P_{1/2}$ state. Ionization producing the $5p^5$ $^2P_{1/2}$ state is only possible through direct non-resonant two-photon excitation above the $5p^5$ $^2P_{1/2}$ threshold. **b** and **c** display measured kinetic energy spectra corresponding to the measured eFROG traces in Fig. 1.**b** and **c**, respectively.

In the case of long pulses with a narrow spectrum (bandwidth < spin-orbit splitting) the two contributions to the photoelectron kinetic energy spectrum do not overlap and a simple spectral filtering can discriminate the two spin-orbit pathways, as was done in a former measurement of 50-fs-long pulses[26]. However, for very short pulses with a broad spectrum spanning more than 1 eV bandwidth as used in our study, the two pathways partially overlap. Therefore, we developed a simple heuristic model assuming that the two ionization channels are independent from each other, and yield two identical contributions with an energy shift corresponding to the spin-orbit splitting. The eFROG trace is then an incoherent sum of both contributions with different weights. Furthermore, the imprint of the auto-ionizing states on the eFROG measurement is taken into account as a contribution to the spectral response function, which is a commonly used approach to match the calculated trace to the measurement in each iteration of the retrieval. We implement this model in the phase retrieval on the basis of a modified differential evolution algorithm (for details see Methods) which runs sufficiently fast so that the evaluations could be performed on a desktop computer.

Before we characterize the tunable VUV pulses, we investigate whether our algorithm based on a simple model can: (i) sufficiently grasp the most significant features of the underlying photoionization processes, and (ii) accurately retrieve the electric field of a short pulse from eFROG traces. First, we perform phase retrievals on the two exemplary measured traces shown in Fig. 1.**b** and **c**. We observe a very good match between the measured (Fig. 3 **a** and **g**) and retrieved (Fig. 3 **b** and **h**) traces suggesting that our model captures well the photoionization mechanisms including the AIS as well as the two ionization channels. The FROG errors on the 74x132 size grid amount to 2.5% and 2.1%, respectively. Furthermore, to address (ii) we test our algorithm against *ab initio* quantum-mechanical calculations based on solving the time-dependent Schrödinger equation (TDSE) (Methods). To that end, the electric fields obtained by the retrievals shown in (Fig. 3 **b** and **h**) are fed into the TDSE solver to calculate the kinetic energy distribution of the photoelectrons at each delay, thus simulating their corresponding eFROG traces. We then use the calculated traces as an input for further phase retrievals (shown in Fig. 3 **d** and **j**). The retrieved pulse shapes (Fig. 3 **e** and **k**) as well as the spectral phases (Fig. 3 **f** and **l**) reproduce very well the known input shapes, which is quantified by the low values of the root-mean-square field errors (ε, see Methods) indicated in Fig. 3 **e** and **k**. These results strongly indicate the suitability of our heuristic model to retrieve the pulse shape from eFROG measurements with high fidelity.

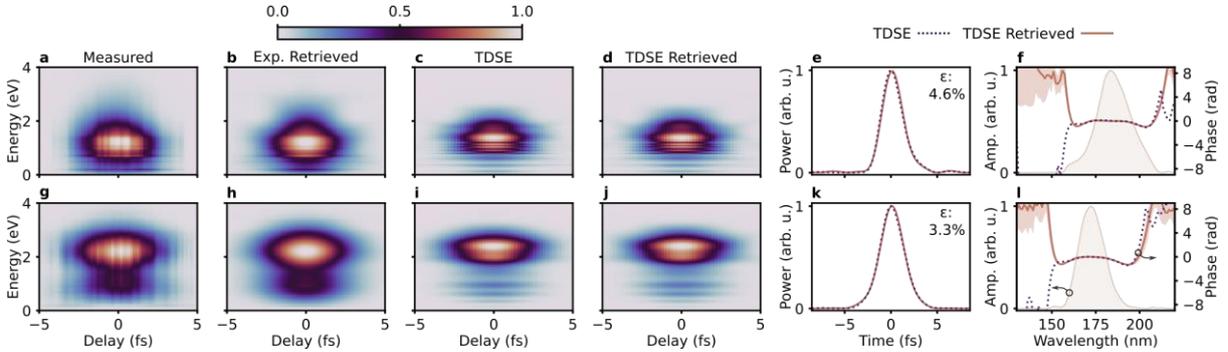

Fig. 3 Test of eFROG for traces exhibiting AIS fingerprints (RDW pulses centered at 180 nm; top row) and contributions of two ionization channels (RDW pulses centered at 170 nm; bottom row). The left column (**a**, **g**) shows the measured, the second (**b**, **h**) the retrieved, and the third column (**c**, **i**) the *ab initio* calculated traces. The fourth column (**d**, **j**) corresponds to retrievals using the simulated traces as an input. The fifth (**e**, **k**) and last (**f**, **l**) columns display the pulse shapes and spectra with spectral phases, respectively. In the last two columns the input shapes of the simulations are denoted by dotted lines, while the retrieved shapes by solid lines.

After gaining confidence in our retrievals, we reconstruct the temporal shape of VUV pulses tuned between 160 and 190 nm from a series of eFROG measurements. The average temporal envelopes and phases retrieved from the eFROG traces using our algorithm are summarized in Fig. 4 (full data in Supplementary Fig. S3). The pulse shapes are very similar for all pulses at different central wavelengths exhibiting a slight chirp. The chirp is attributed partly to the generation process, partly to the propagation through the dilute gas medium at a few mbar pressure originating from the capillary. The pulse durations defined by the full width at half maximum (FWHM) are below 3 fs with the exception of the pulse with the central wavelength of 190 nm (pulse D in Fig. 4). We note that in the latter case the slightly increased pulse duration is accompanied by a doubling of the pulse energy yielding a maximal peak power of ~0.8 GW.

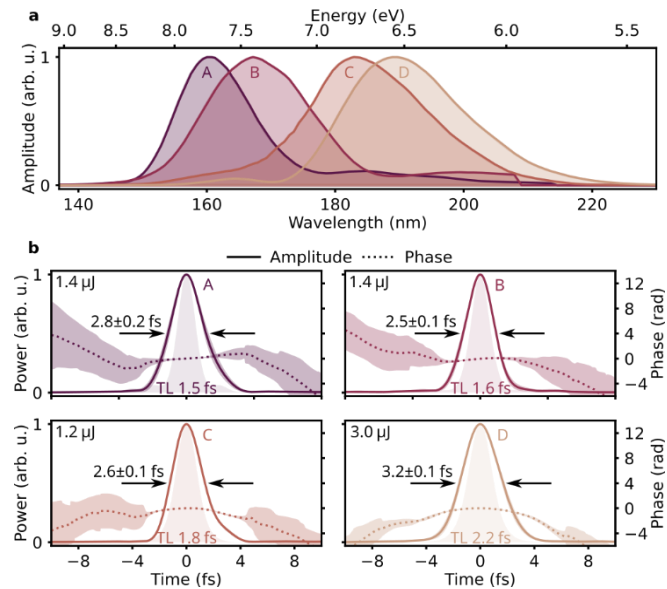

Fig. 4 Fully characterized VUV pulses tuned between 160 and 190 nm (marked by A to D). **a**, The spectra measured by a VUV spectrometer. **b**, The average temporal envelopes (solid) and phases (dotted) retrieved from 10 independent retrievals of the measured eFROG traces. The shaded areas surrounding the envelopes and phases correspond to the standard deviation of the retrievals. The transform limited (TL) pulse shapes corresponding to the measured spectra are displayed as shaded backgrounds. The full width at half maximum (FWHM) values of each shape are also included. The measured VUV pulse energies are indicated in the upper left corner of each panel.

**Discussion**

We compare the measured pulse shapes with the results of simulations proven to describe the generation of RDW pulses at high fidelity. For the simulations we solve the unidirectional pulse propagation equation along the hollow waveguide taking the first four eigenmodes into account using the open-source "Luna" library (https://github.com/LupoLab/Luna.jl). The details of the simulations including the full set of input parameters corresponding to our VUV generation conditions are summarized in the Supplementary S5.

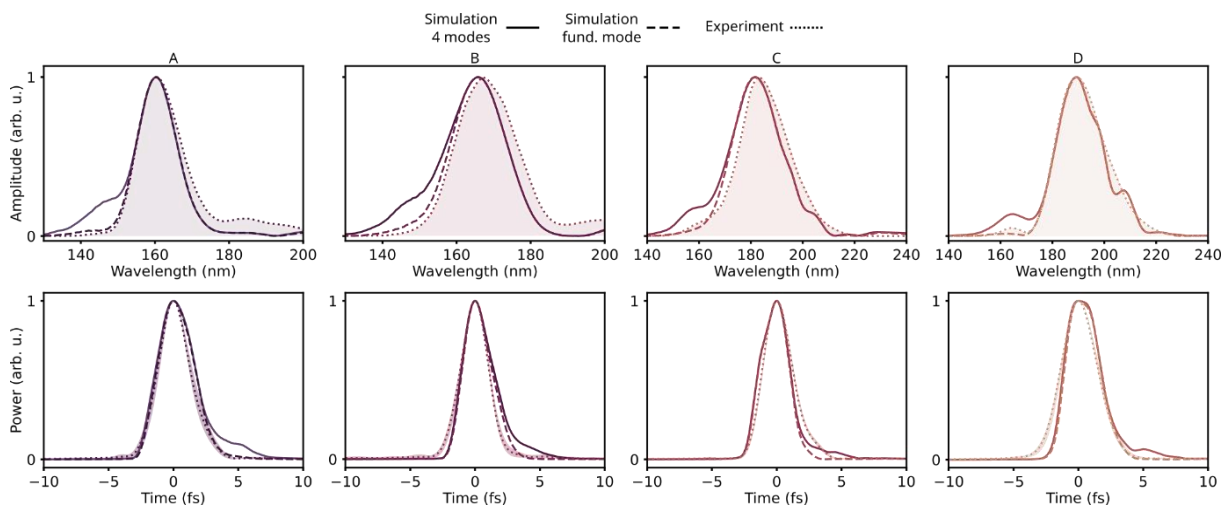

Fig. 5 Comparison between RDW pulse measurements and simulations taking the first four eigenmodes of the waveguide into account. The columns correspond to the pulses A, B, C and D of Fig 3. The spectra are displayed in the top row while the pulse shapes are shown in the bottom row. Solid lines represent the output of the simulation, the dashed lines correspond to the fundamental eigenmode, and the dotted lines denote the experimentally measured data. In the top row the shaded areas correspond to the measured spectra, while in the bottom row the shaded regions mark the standard deviation of the retrieved pulse shapes.

The results of the simulation are compared with the measured spectra and pulse shapes in Fig. 5 exhibiting an excellent overall agreement. Interestingly, the agreement is better between the filtered fundamental mode and the measurement, which we attribute to the filtering effect of the finite aperture of the steering mirrors during the long propagation towards the eFROG apparatus.

As an experimental validation of our method, we determine the phase shift due to an ultra-thin fused silica plate placed into the VUV beam. We measure eFROG traces with and without inserting a 10 μm-thin UV fused silica plate in the beam. The measured phase shift introduced by the plate is the difference between the two retrieved phases (solid curve in Fig. 6). For comparison, the phase shift due to the plate (dashed curve) is calculated using the VUV refractive index of fused silica[29] in combination with the thickness of the plate (Methods).

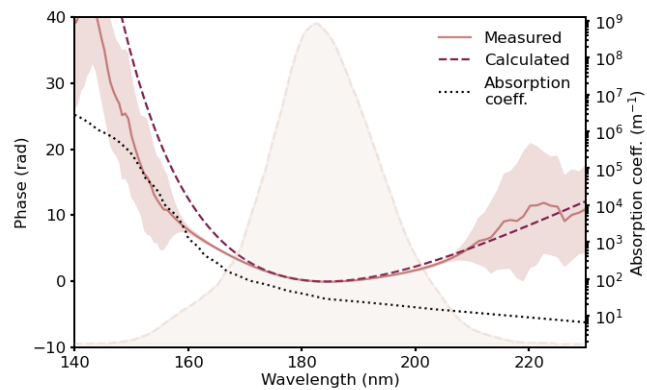

Fig. 6. Phase shift due to a 10-μm-thin fused silica plate. The solid line corresponds to the difference between the average phases of 10 retrievals with and without the plate. The shaded area marks the uncertainty of the measured phase difference. The calculated phase imparted by the silica plate is plotted by the dotted dark red line. The large absorption coefficient of fused silica is displayed as a black dotted line. The spectrum of the VUV pulse is shown in the background for reference.

The agreement between the measured and the calculated phase difference is remarkable. Only a slight deviation of the two curves is observable below 170 nm, where the absorption of fused silica rapidly increases distortions of the pulse.

**Conclusion**

Until recently, the lack of few-cycle UV pulses prevented fundamental studies in a spectral range, where most materials exhibit electronic resonances. High-energy RDW emission emerging from a hollow capillary fiber[21] offers a versatile light source for such experiments. Indeed, the first pump-probe experiments involving few-fs DUV pulses were performed recently[30-32], however, no studies have been reported so far in the more challenging VUV spectral range. Although numerical simulations suggest that VUV pulses due to RDW emission can also be very short (few-fs duration), up to now no experimental evidence proved these predictions. Here we demonstrate the generation of VUV pulses tunable between 160 and 190 nm with sub-3 fs duration carrying a few μJ of energy. For the in-situ pulse characterization we implemented a photoionization-based FROG technique, which is generally applicable to short VUV pulses down to few-fs. The controlled generation of well-characterized few-cycle VUV pulses unlocks the study of coupled electronic and nuclear motion in electronically excited molecules with an unprecedented time resolution. Our approach is capable of performing pump-probe experiments in the VUV with few-fs resolution. A series of such experiments on small molecules are currently underway in our laboratory. They will shed new light on early-time dynamics in photo-excited molecules capturing the interest of the scientific community.

**Methods**

*VUV generation*

We send 37 fs pulses with 2 mJ energy of a Ti:sapphire chirped pulse amplifier system (Spectra Physics Spitfire Pro) operating at a repetition rate of 1 kHz into a 3-m-long SF-HCF (HCF1 in Fig. S1) with a core diameter of 450 µm. The capillary is differentially filled with argon gas at a backing pressure of 120 mbar. The spectrally broadened pulses are subsequently compressed to 10 fs duration by a set of chirped mirrors (PC70, Ultrafast Innovations). The light is then re-focused into a second, 1.5-m-long SF-HCF of 250 µm core diameter (HCF2 in Fig. S1) where soliton self-compression and RDW emission take place. The pulse energy entering the second capillary can be set by a variable attenuator consisting of a half-wave plate at the output of the laser and a broadband wire-grid polarizer placed between the capillaries before the chirped mirrors. The maximum available pulse energy for inducing soliton dynamics is 680 µJ and the transmission of the evacuated SF-HCF is ~60%. We used HCF2 in an inverse pressure gradient mode, where we apply helium up to 2 bar pressure at the entrance of the waveguide, while the output is directly connected to a vacuum beamline. After collimation of the beam using an aluminium-coated concave mirror, the broadband soliton content was filtered out using two Brewster-angled silicon plates at an angle of incidence of 75 deg.

*VUV spectrum and power measurement*

Two motorized mirrors allow to send the beam to different diagnostics. The spectrum was measured using a McPherson 234/302 spectrograph equipped with 1200 lines/mm Al+MgF$_2$-coated flat-field grating and a CCD array (Andor DU920N). The spectrometer was calibrated using an Hg-Ar spectral line lamp. The power is measured by a calibrated volume absorber power meter (Ophir 3A-P) placed directly in the vacuum chamber. We measured the background signal remaining from the broadband soliton radiation due to incomplete filtering by applying the full available laser power through HCF2 and adjusting the gas pressure to induce soliton self-compression but remain right below the threshold of dispersive wave generation. In this way we measure 40 µW (corresponding to 40 nJ pulse energy), close to the power detection minimum.

*The VUV eFROG apparatus*

The VUV beam was split into two halves by two D-shaped aluminium mirrors, one of which was mounted on a piezo-driven nano-positioning stage (P-625.1U, Physik Instrumente GmbH). This allows us to have two time-shifted replicas of the VUV pulses with controllable delay. These were routed and focused into a chamber housing a VMI unit, with the focus at the center of the apparatus. The steady-state gas pressure in the VMI during measurement was 2~5x10$^{-5}$ mbar. Velocity maps were obtained by a microchannel plate and a phosphor screen for photoelectron visualization and a CCD camera for recording the images. Scans of the delay between the two-replicas were performed while recording the VMI signal. The procedure of extracting a FROG trace from the pictures is detailed in the supplement.

*Phase retrieval from the eFROG traces*

For the phase retrieval of the pulses from the measured eFROG traces a differential evolution code was implemented[33]. In order to account for the effect of the spin-orbit splitting, the retrieved trace was calculated as follows:

First the contributions of each spin-orbit channel are calculated:

$$S_{1\tau}(t) = F(t) \cdot F(t-\tau) \cdot \exp(-i\omega_1 t) \text{ and } S_{2\tau}(t) = F(t) \cdot F(t-\tau) \cdot \exp(-i\omega_2 t)$$

where $\omega_i = IP_i/\hbar$ (i=1, 2) are the optical frequencies corresponding to each of the ionization potentials and $F(t)$ denotes the complex electric field of the pulse.

Then the kinetic energy spectrum of the photoelectrons at a given time delay τ is calculated by the incoherent sum of both contributions:

$$T_{1\tau}(\omega) = |F\{S_{1\tau}(t)\}|^2 \text{ and } T_{2\tau}(\omega) = |F\{S_{2\tau}(t)\}|^2$$

$$T_\tau(E) = T_{1\tau}(E/\hbar) + C \cdot T_{1\tau}(E/\hbar)$$

where $F$[28] denotes the Fourier transform, $E$ the kinetic energy of the photoelectrons and $C$ is the ratio of the yields of both channels. The value of $C$ is also one of the parameters being retrieved as it is included in the properties of each individual of the differential evolution population. Finally, the trace is corrected by a delay-independent "response function" $\mu(E)$ which minimizes the difference between the two traces[34]:

$$\mu(E_i) = \frac{\sum_j T_{meas}(E_i, \tau_j) T(E_i, \tau_j)}{\sum_j T(E_i, \tau_j)^2}.$$

This is necessary to correctly include the spectral sensitivity of the apparatus and the frequency dependence of the two-photon ionization process including the contributions of the auto-ionizing states.

The spectral phase is represented by 6 coefficients of its Taylor expansion, from the second order (GDD) term and above. These coefficients are retrieved by the differential evolution algorithm. Each run starts with a population of 20 individuals all with random coefficients. For each individual a trace was calculated and compared to the measured one by computing the RMS of the residuals[34]. This value was then used to rank the individual within the population. For all the presented data, 10 retrievals were performed for each measured eFROG trace for statistics.

In Fig. 3 **e** and **k** we compare the input and retrieved pulse shapes. For a quantitative comparison we calculate the rms field error[35]

$$\varepsilon = \sqrt{\frac{1}{N} \sum_{n=1}^{N} |F_n^0 - F_n|^2 / \max |F^0|^2},$$

where $F^0$ is the known input, and $F$ is the retrieved electric field, both sampled by $N$ points. Before calculation the amplitudes, the carrier envelope phases and the group delays of the electric fields are matched[34].

*Quantum mechanical code for calculating the kinetic energy spectrum of the photoelectrons*

The time-dependent Schrödinger equation (TDSE) calculations were performed within the time-dependent configuration-interaction singles (TD-CIS) *Ansatz*, wherein single excitations are allowed from a reference state, in the present case from the Hartree–Fock ground state. Our implementation of TD-CIS is described in [36,37]. To account for quasirelativistic effects, we employ a small-core relativistic effective-core potential (RECP) designed by Peterson et al.[38], which yields correct spin–orbit splitting but slightly too low ionization potential (see Table 1). The calculations were performed on a spherical grid, where the radial grid extended to 300 Bohr, and the grid points were distributed according to

$$r_j = r_{j-1} + \rho_{min} + (1 - e^{-\alpha r_{j-1}})(\rho_{max} - \rho_{min}), r_1 = \rho_{min}/2,$$

and $\rho_{min}$, $\rho_{max}$, and $\alpha$ were chosen to be 0.12 Bohr, 0.17 Bohr, and 0.3, respectively. The spin–angular dimensions were expanded using all spinor spherical harmonics (See 7.2 of Varshalovich[39]) for which the orbital angular momentum $l \leq 10$. Since we consider linear polarization only, we may use the restriction $\Delta m_j = 0$, for a total of 164 spin–angular partial waves, since we additionally limited excitation to occur only from ns² np⁶. To avoid reflections at the edge of the computational box, we employ the complex-absorbing potential by Manolopoulos[40], spanning the last 231.77 Bohr of the box, and with a design parameter δ=0.20866.

| Element | Orbital | Theory (eV) | Experiment (eV) | Δ (eV) | Rel. Δ (eV) |
|---|---|---|---|---|---|
| Xe | $5p_{3/2}$ | 12.026 | 12.130 | -0.104 | -0.9 |
|  | $5p_{1/2}$ | 13.483 | 13.436 | 0.047 | 0.3 |
|  | $5s_{1/2}$ | 27.927 | 23.397 | 4.530 | 19.4 |

TABLE 1 Theoretical ionization potentials of the ns and np electrons of xenon, compared with their experimental values (taken from[41,42]).

The retrieval algorithm yields the electric field amplitude discretized on a uniform grid: $\{t_n, F_c(t_n)\}$, where $t_n = t_0 + n\delta t$ and $F_c(t_n) \in \mathbb{C}$ are complex due to the spectral nature of the algorithm. The TDSE calculations need the vector potential $A(t)$ [which is related to the electric field as $F(t) = -\partial_t A(t)$], and we compute this by integration in the Fourier domain. Additionally a finer time grid than provided by the retrieval algorithm is necessary (we take 10 steps/au ≈ 413 steps/fs), and we employ cubic Hermite interpolation for that purpose. Finally, the electric field amplitude and vector potentials need to be real quantities, so we arbitrarily choose the real part of the interpolated functions: $F(t) = \Re\{F_c(t)\}, A(t) = \Re\{A_c(t)\}$.

The retrieved pulse shape and a delayed copy are fed as input to the TDSE calculations, and the resulting photoelectron spectrum is computed using the surface-flux techniques[36,43,44]. This calculation is repeated for a set of delays in range −5 fs to 5 fs. Since for even very large pump–probe delays, the two-photon ionization signal is non-negligible, we subtract a "background" taken as the photoelectron spectrum resulting from a 20 fs pump–probe delay calculation. To approximately account for the phase slip across the spatial profile of the pulse due to the noncollinear geometry, we repeat the delay scan, applying a carrier–envelope phase (CEP) shift to the delayed pulse replica. The fields used in the calculation are thus

$$F(t) = \Re\{F_c(t)\} + cos\theta\Re\{F_c(t-\tau)\} + sin\theta\Im\{F_c(t-\tau)\},$$
$$A(t) = \Re\{A_c(t)\} + cos\theta\Re\{A_c(t-\tau)\} + sin\theta\Im\{A_c(t-\tau)\},$$

where $\tau$ is the time delay and $\theta$ the applied CEP. The resultant delay scans for four different CEP values are shown below, as well as their average.

*Measurement of the thickness of the fused silica plate*

For the calculation of the chirp introduced by a fused silica plate the knowledge of the sample thickness is necessary. As the plate is only a few μm thick, we used spectral interferometry to precisely determine its thickness. The spacing of the spectral fringes arising from the interference between the front and back reflections is measured when the plate was illuminated by the Ti:sapphire laser oscillator.

**Acknowledgements**

We thank Oleg Kornilov (MBI) and Sven Kleinert (Leibniz Universität Hannover) for fruitful discussions concerning the vacuum beamline and the phase retrieval algorithm, respectively.

This work was funded by the Deutsche Forschungsgemeinschaft (DFG) (NA1102/3-1). A.R acknowledges support by the Leibniz Gemeinschaft (SAW-K380-2021).

**Author contributions**

T.N., A.M.B. and A.R. conceived this work. J.R.C.A. and T.N. built the VUV source. M.K. proposed the eFROG method. J.R.C.A., M.K. and R.D. with the help of A.R. performed the eFROG measurements. J.R.C.A., M.K. with the help of T.W. processed the data. J.R.C.A. wrote the phase retrieval algorithm and evaluated the eFROG traces. S.C. with the help of S.P. performed the TDSE simulations. M.J.J.V., M.Yu.I., A.R. and T.N. interpreted and analysed the results. T.N., A.R., J.R.C.A. and M.K. drafted the manuscript and all authors participated in the discussion of the results, and the editing the manuscript.

**Competing interests**

The authors declare no competing interests.

# Supplementary material

## S1. Experimental arrangement

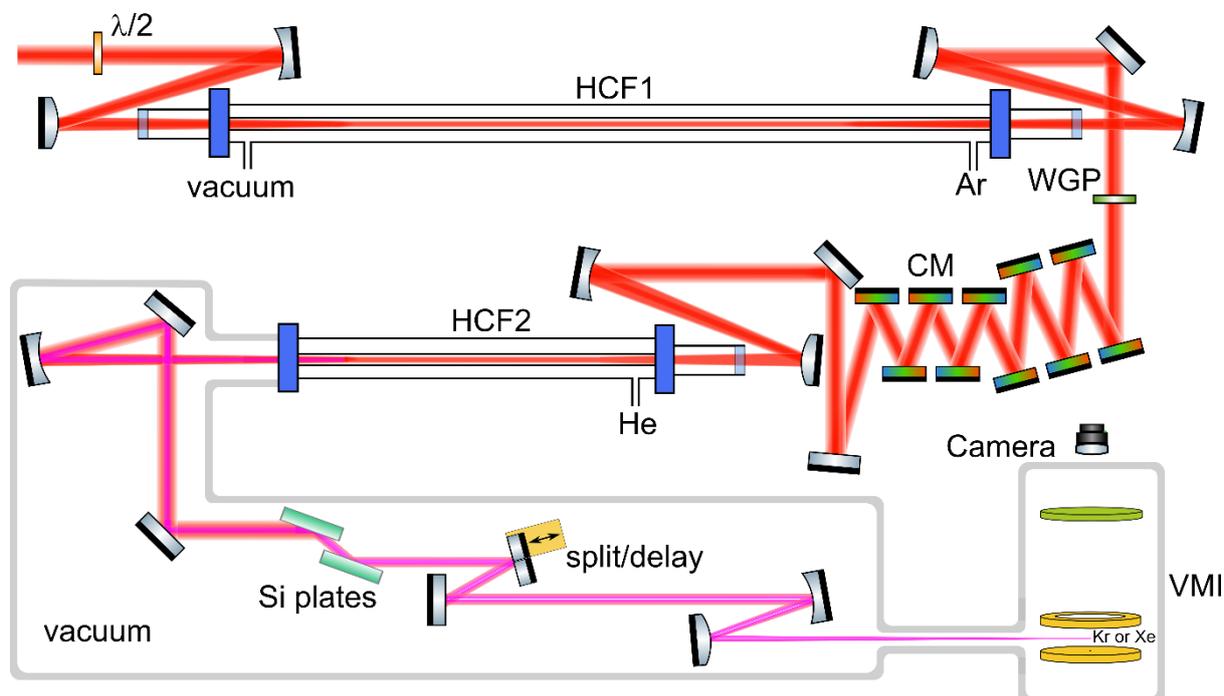

FIG. S1. Detailed experimental arrangement. λ/2 denotes a half-wave plate, which forms a variable attenuator together with a wire-grid polarizer (WGP). CM stands for a set of chirped mirrors, HCF for hollow-core fiber and VMI for velocity-map imaging spectrometer.

## S2. Choice of noble gases for VUV eFROG

| Gas | ½ Ip (eV) | Lowest Rydberg state (eV) | Wavelength range (nm) |
|---|---|---|---|
| helium | 12.30 | 19.82 | 62-100 |
| neon | 10.78 | 16.67 | 74-115 |
| argon | 7.88 | 11.62 | 107-157 |
| krypton | 7.00 | 10.03 | 124-177 |
| xenon | 6.07 | 8.44 | 147-204 |

TABLE S1. Gases for non-sequential two-photon ionization by VUV pulses

## S3. Data acquisition and processing

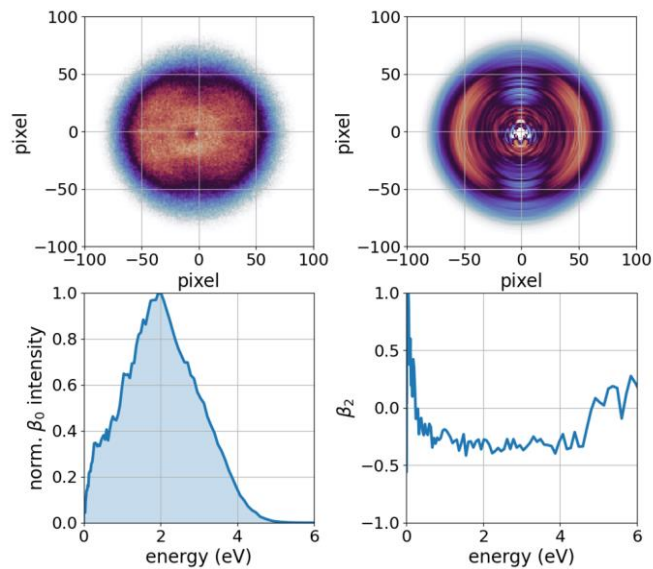

FIG. S2. The recorded VMI-image (a), is Abel-inverted to yield (b) the central velocity map of photoelectrons. The signal is integrated over the two spherical angles and energy-calibrated to obtain (c) the photoelectron spectrum corresponding to the lowest order Legendre-decomposition $\beta_0$. Higher orders, shown exemplarily in (d) for $\beta_2$, can further deduce on the signal anisotropy.

In the experiment, a velocity-map imaging spectrometer (VMI) is used to detect the kinetic energy spectrum of the photoelectrons originating from two-photon ionization. The data acquisition and data-processing procedure are summarized in Fig. S2. The photoelectrons are imaged by an electrostatic lens arrangement onto the surface of a multi-channel plate. It spatially maps a projection of the electron velocity distribution on a two-dimensional VMI-image, shown in Fig.S2 (a). The two-dimensional image is then Abel-inverted in order to obtain a central slice through the reconstructed three-dimensional Newton-sphere. The Abel-inversion is performed by the rBasex method implemented in the PyAbel code (https://pypi.org/project/PyAbel), which is also used for further data processing. The retrieved central slice, shown in Fig. S2 (b), represents the velocity distribution of the photoelectrons undergoing photoionization. The photoelectron energy axis is calibrated by the auto-ionising states present in the traces and then corroborated by recording above-threshold-ionization (ATI) peaks from photoionization of xenon and krypton by narrowband 30 fs, 800 nm driving laser pulses. The radial intensity distribution calculated by integrating over both spherical angles of the Newton-sphere, shown in Fig S2 (c), is the $0^{th}$-order component of the Legendre decomposition of the Abel-inverted image which is denoted by $\beta_0$. After Jacobian correction it represents the photoelectron spectrum retrieved from the measured two-dimensional VMI image. Furthermore, the higher-order components of the Legendre decomposition, such as the $\beta_2$ component shown in Fig. S2 (d), provide additional information on the anisotropy of the detected photoelectron signal.

In order to record an eFROG trace the delay between the two VUV-pulse replicas was scanned from -13 fs to +13 fs with a step size of 0.13 fs. For each delay a large number of shots (800-2400, dependent on the signal level) are averaged to yield a VMI-image with good signal-to-noise properties. The scans are repeated five to ten times and summed up to form a low-noise eFROG trace. As only one of the pulses already can ionize the gas, the trace exhibits a constant background along the delay axis. We determine the background kinetic energy spectrum by averaging those at both large positive and negative delays, and subtract this from the eFROG trace.

## S4. Full data of the eFROG measurements

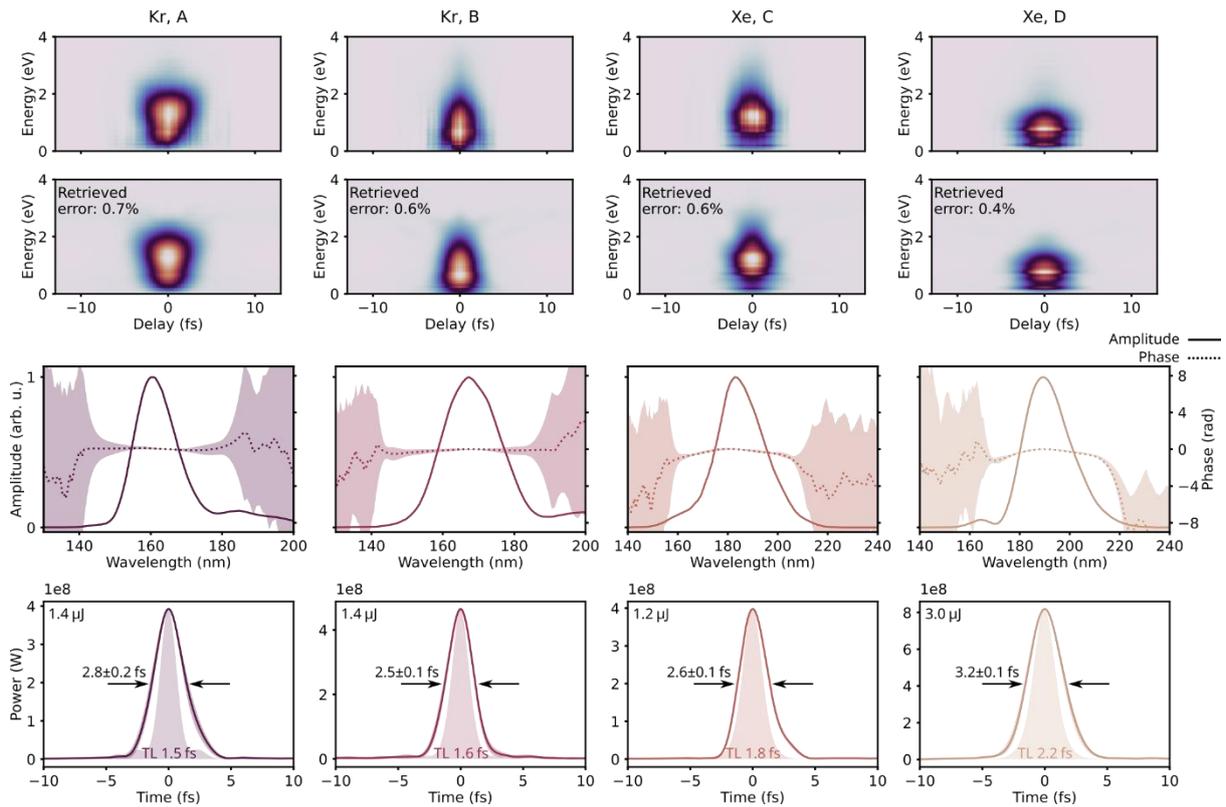

FIG. S3. Full eFROG measurements corresponding to the results presented in Fig. 4. Each column contains a full measurement of the pulses A, B, C or D of Fig. 4. The rows from top to bottom show the measured and retrieved eFROG traces, the retrieved spectral and temporal shapes for each pulse. The shaded areas around the curves denote the standard deviation of the spectral phase and the pulse shape in the third and fourth row, respectively. The shaded backgrounds in the bottom row indicate the transform-limited pulse shapes. The retrieved traces also include the FROG errors for the grid of 196x1024.

## S5. Simulations of the RDW generation

The input electric field of the RDW generation process taking place in HCF2 is determined by a series of 6 consecutive SHG-FROG measurements at gradually increasing wedge insertions. This was necessary to unambiguously determine the direction of the time axis, which is impossible in case of a single SHG-FROG measurement. The phase was corrected for taking the differences in the beam path between the FROG measurement and the RDW generation into account. The input pulse shape is shown in Fig. S4.

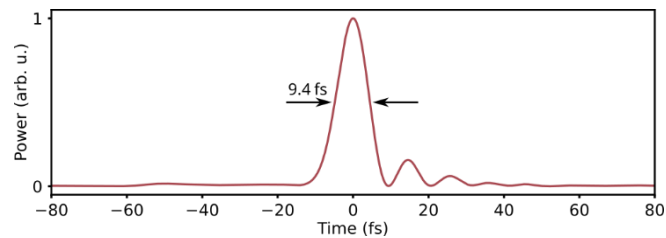

FIG. S4 Measured input pulse for RDW generation.

We performed multi-mode simulations including the first four hybrid modes[45] ($EH_{11}$ to $EH_{14}$). In Fig. 5 we also show the content of the fundamental mode where we filtered the $EH_{11}$ mode from the results of the multi-mode propagation. The simulations were done in inverse pressure gradient mode starting from the pressures indicated in Table S2 to a pressure of 4 mbar with a fiber length of 1.5 m and an inner diameter of 250 μm. In cases of B and D the pressure was slightly adjusted to avoid a substantial discrepancy in the resulting central wavelength of the RDW.

|   | Pressure (bar) | | Input energy (μJ) | Pulse duration FWHM (fs) | | |
| --- | --- | --- | --- | --- | --- | --- |
|   | Experiment | Simulation | Experiment | Experiment | Simulation all 4 modes | Simulation only fund. mode |
| A | 1.35 | 1.35 | 495 | 2.8 | 3.4 | 3.3 |
| B | 1.5 | 1.55 | 385 | 2.5 | 2.8 | 2.6 |
| C | 1.9 | 1.9 | 314 | 2.6 | 2.7 | 2.6 |
| D | 1.9 | 2 | 330 | 3.2 | 3.1 | 3.1 |

TABLE S2: Experimental values and input parameters/results of the fiber propagation simulations.

The experimentally measured input energies were scaled down to 60% level in order to take the in-coupling losses into account. The pulses are obtained by spectrally filtering the RDW from the driving pulses.

### S6. Comparison of gas targets in eFROG measurements

Here we also show that the specific choice of the target gas species (allowed by Table S1) does not influence the eFROG measurement results. For this purpose, we measure RDW pulses centered at 170 nm using both krypton and xenon as target medium. Although the structure of the eFROG traces shown in Fig. S5 are substantially different, the retrievals yield in both cases essentially the same pulse shape. We also note that the two measurements were performed with a time difference of two hours, which is a further indication of the excellent stability and reproducibility of both the VUV pulse generation and characterization.

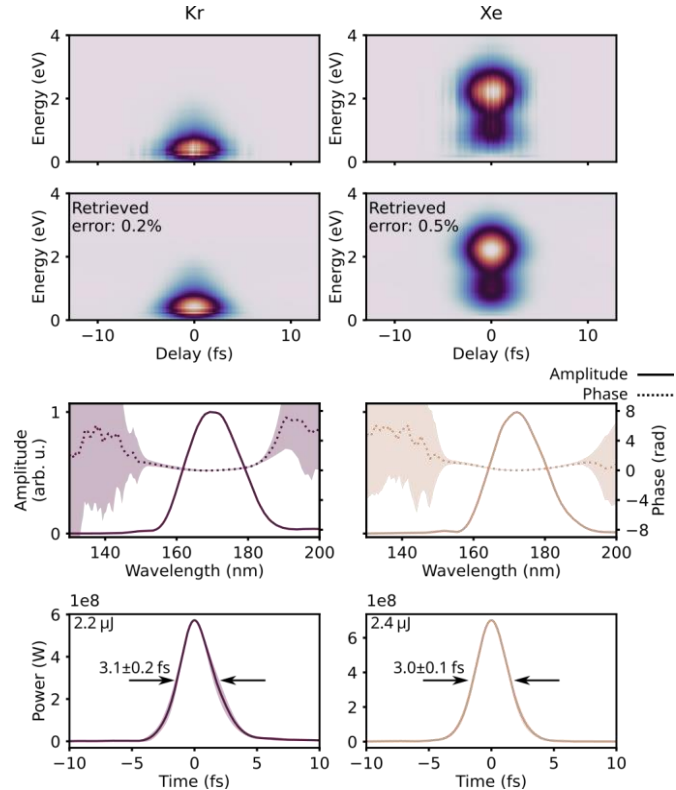

Fig. S5. eFROG measurements of the same VUV pulse using Kr (left column) and Xe (right column) as target gases. Top row: measured eFROG traces; second row: retrieved traces; third row: retrieved spectral intensities and phases; bottom row: retrieved pulse shapes. Both measurements yield the same pulse within the experimental uncertainties.